# Improving Geopolitical Forecasts with Bayesian Networks

Matthew Martin[a]

[a]Université du Québec à Montréal

Montreal, Quebec, Canada

Correspondence: martin.matthew@courrier.uqam.ca

201, avenue du Président-Kennedy

Local PK 4150

H2X 3Y7 Montréal (Québec)

**Acknowledgements**

The author wishes to thank Serge Robert, Roger Villemaire, and Stephen Martin for their insightful comments and help in editing the final manuscript.

**Competing Interests Statement**

Competing interests: The author declares none.

**Financial Support Statement**

This work was supported by the Fonds de recherche du Québec – Nature et technologies [grant number 303239].

**Publication Status**

Submitted to the International Journal of Forecasting on January 6, 2026.

**Abstract**

This study explores how Bayesian networks (BNs) can improve forecast accuracy compared to logistic regression and recalibration and aggregation methods, using data from the Good Judgment Project. Regularized logistic regression models and a baseline recalibrated aggregate were compared to two types of BNs: structure-learned BNs with arcs between predictors, and naive BNs. Four predictor variables were examined: absolute difference from the aggregate, forecast value, days prior to question close, and mean standardized Brier score. Results indicated the recalibrated aggregate achieved the highest accuracy (AUC = 0.985), followed by both types of BNs, then the logistic regression models. Performance of the BNs was likely harmed by reduced information from the discretization process and violation of the assumption of linearity likely harmed the logistic regression models. Future research should explore hybrid approaches combining BNs with logistic regression, examine additional predictor variables, and account for hierarchical data dependencies.

# 1. Introduction

Governments, militaries, businesses, and other stakeholders must make decisions and long-term plans in the face of uncertainty about the future state of the economy, the actions of other states, and world events. Anticipating important geopolitical developments allows stakeholders to make and execute strategic plans to better achieve their goals. Increasing the accuracy of geopolitical forecasts thus has immense value to these actors. To this end, the US government has funded several projects related to improving geopolitical forecasting, including the Aggregative Contingent Estimation (ACE) tournament in 2010, in which teams competed to find the best means of eliciting, aggregating, and representing probabilistic judgements of future geopolitical events (IARPA, 2010). The winning team was the Good Judgement Project (GJP) led by Phil Tetlock and Barbara Mellers, who used participant training, teaming, and forecast recalibration and aggregation techniques to achieve accuracy superior to professional intelligence analysts who had access to classified information (Goldstein et al., 2015; Good Judgment Inc., 2025).

Since the ACE tournament, national governments have begun using similar forecasting techniques on an ongoing basis, such as the UK's Cosmic Bazaar (The Economist, 2021) and the RAND forecasting initiative, which advises the US government (RAND, 2025). Businesses have also begun to use forecasting techniques to predict the outcomes of their internal projects and competitive environments, such as Twitch and Google (Hernandez, 2017; Schwarz, 2024). Organizations such as Metaculus and Good Judgement Open also provide forecasting platforms that allow questions about a diverse range of topics for a wide range of clients (Good Judgment Open, 2025; Metaculus, 2025).

Researchers have discovered several psychological variables predictive of forecasting accuracy. According to Mellers et al. (2015), the most important are measures related to a forecaster's behaviours such as the number of forecasts per question and the amount of time spent deliberating per question. Situational variables such as training in probabilistic reasoning and being a member of a team are also important factors predicting accuracy, as are dispositional variables such as a general intelligence, political knowledge, and open-minded thinking (Mellers et al., 2015).

Atanasov and Himmelstein (2023) conducted a separate review of the predictors of forecasting accuracy. In summary, the authors suggest that the metrics most associated with an individual's success in predicting outcomes are, in order of descending value: accuracy-related (e.g., accuracy on previously resolved questions), intersubjective (e.g., squared distance from a consensus forecast), behavioural (e.g., frequency of updating), dispositional (e.g., intelligence and personality), and then finally expertise-related (e.g., knowledge tests). The authors suggest that these metrics would serve as useful means of giving higher weight to forecasters that exhibit these characteristics.

Relatedly, researchers have explored various ways of using algorithmic and statistical techniques to enhance the information available from raw forecasted probabilities to predict geopolitical events. These techniques range from statistical techniques like regression (Cross et al., 2018), to aggregation and recalibration (Atanasov et al., 2017), and machine learning (Shinitzky et al., 2024). To this end, Shinitzky et al. (2024) explored the extent to which standard machine learning techniques can improve the accuracy of geopolitical forecasts. They tested regularized versions of Bernoulli naive Bayes, gradient boosting classifier, linear discriminant analysis, and logistic regression on their ability to predict binary event outcomes given participants' forecast probabilities and over 40 other predictor variables.

The additional predictor variables included variables derived from the group's forecasted probabilities (e.g., the absolute difference of a forecast from the mean), other features elicited from the forecasters (such as which dimensions they considered in making their prediction), linguistic features from the forecast rationales, features about the participants' psychology (e.g., locus of control) and demographics (e.g., education level), as well as ratings provided by external evaluators about the forecasts' convincingness. Shinitzky et al. (2024) found that regularized logistic regression had the highest performance of all machine learning techniques they tested, as measured by area under the receiver operating characteristic curve (AUC). Using regularized logistic regression, they then tested combinations of their features to see which were most predictive of the outcomes of geopolitical events. They found that the extremeness of the forecasted probability and absolute distance from the group consensus were the most important variables, followed by linguistic features of the forecast rationales. Finally, they used the classification probabilities from the winning logistic regression model as weights in an aggregation procedure, showing that the probability weights increased the accuracy of an aggregate prediction, compared to a simple confidence-weighted baseline aggregate.

While Shinitzky et al. (2024) helpfully explored how standard machine learning techniques can be applied to improve forecasting accuracy, questions can be raised about how replicable their results would be in contexts more commonly found on forecasting platforms such as Good Judgement Open or Metaculus. The first consideration is that their sample was relatively small, consisting of only 153 participants and 10 questions. For comparison, the original ACE GJP tournament consisted of over 9000 participants and 600 questions. Part of the rationale behind prediction platforms is to leverage the wisdom of the crowd, and thus maximizing the number of forecasters, questions, and answers per question is an important goal. A second consideration is that the format of Shinitzky et al.'s (2024) tournament did not use accuracy metrics as predictors in their main analysis, as no questions had resolved at the time of forecast elicitation. While there are certainly situations where such 'cold-start' conditions hold, accuracy-related scores are the best class of predictors for future accuracy and are an important part of forecast aggregation procedures (Atanasov et al., 2017; Atanasov & Himmelstein, 2023). A third consideration is that for 9/10 of the questions, only one forecast per participant was elicited, with only one question allowing for a single update. This is problematic because some of the non-accuracy-related metrics most predictive of accuracy are the frequency and magnitude of updates to forecasts (Atanasov & Himmelstein, 2023). A fourth consideration is that the time range of Shinitzky et al.'s (2024) forecasting period was relatively restricted, where predicted events were between one and four months in the future. In comparison, the GJP tournament had forecasts ranging from 1 to 549 days in the future. In longer term forecasting situations, the change in estimated probability over time is often decision-relevant, as it can signal important geopolitical developments necessitating strategic choices. Finally, some of the metrics used as predictors of outcomes in Shinitzky et al.'s (2024) study would be impractical to implement in large-scale settings, such as third-party evaluations of the convincingness of other forecasters' rationales.

Given the importance of improving geopolitical forecasting, researchers should examine alternative approaches and whether Shinitzky et al.'s (2024) results generalize to forecasting contexts more similar to those used by governments, where there is a wider variety and number of participants, forecasts per participant, and timing of forecasts. These contexts also allow for examining forecaster accuracy and other behaviours such as update frequency. To this end, the current study uses data from the original GJP ACE tournament (Good Judgment Project, 2016), which has been studied extensively and serves as an excellent source of crowdsourced forecasting tournament data akin to the other platforms cited above (Ungar et al., 2012).

While regularized logistic regression was the most effective machine learning technique studied by Shinitzky et al. (2024), its assumptions may not hold in a dataset with a greater variety and number of participants, forecasts, and variables unavailable in the 'cold start' situation Shinitzky et al. considered. Additionally, Shinitzky et al. tested a Bernouli naïve Bayes classifier, but did not examine Bayesian networks (BNs) with more complex conditional dependencies between predictor variables. Bayesian networks make fewer parametric assumptions than logistic regression and their ability to represent complex conditional dependencies between variables may prove beneficial in modeling more complex forecasting data. Finally, Shinitzky et al. used an unweighted mean as their aggregation procedure in their baseline predictor variable. More sophisticated aggregation procedures, such as those that incorporate recalibration and weighting by participant characteristics would likely make the models more predictive. Recalibrated weighted forecasts are also highly accurate predictors on their own, so it would be insightful to compare other modeling approaches to these techniques (Atanasov et al., 2017).

## 1.1. Bayesian Networks to Predict Geopolitical Outcomes

The current study explores Bayesian networks (BNs) as an alternative machine learning approach to improving the predictive power of geopolitical forecasts and compares their relative merits to logistic regression and recalibrated aggregate forecasts. Bayesian networks represent relationships between variables as a graph where connections (also called 'arcs') between variables (also called 'nodes') represent conditional probabilistic dependencies. Given that the connections represent conditional probabilities, for a BN to be fully specified, all connections must be directed (specifying which variable is conditional on which), and there can be no cyclical connections. Thus, BNs are directed acyclic graphs (DAGs). The absence of an arc between nodes represents conditional independence. The conditional probability distribution of a variable is determined by its parent nodes, and a variable is conditionally independent of its non-descendants, given its parents. Arcs can be learned via algorithmic means, via manual specification, or a combination of the two (Kitson et al., 2023).

Three common types of BNs are: 1) discrete BNs, where all variables are categorical and conditional relationships are represented in conditional probability tables, 2) Gaussian BNs, which represent all variables as normally distributed continuous with conditional relationships usually expressed as linear regression equations; and 3) hybrid BNs, which include both discrete and continuous nodes, usually with the constraint that discrete nodes cannot inherit from continuous nodes (Kitson et al., 2023). In the current study, only discrete BNs are used since the binary outcome variable could not inherit from continuous variables. A process of discretizing continuous variables is thus employed to meet the constraints of discrete BNs.

Bayesian networks allow one to model complex conditional dependencies between variables, where the relationships between descendent variables depend on the states of ancestor variables. Learning the pattern of conditional dependencies between predictor variables may improve the overall accuracy of the BNs. For example, divergence from a consensus forecast is usually related to poor performance. However, in some cases where a divergent forecaster is otherwise accurate, their divergent forecast should be given more weight, as it represents unique information (Atanasov & Himmelstein, 2023). Thus, the relationship between a divergent forecast and the outcome may be different depending on the states of other variables. To examine the effect of complex conditional dependencies on the BNs' performance, complex BNs whose structures are learned primarily through algorithmic means (henceforth 'structure-learned BNs') are compared to naïve flat-structured BNs (henceforth 'flat BNs') that assume independence between all predictor variables.

Discrete BNs make fewer parametric assumptions about the data than do logistic regression models, which assume, among other things, that there are no influential outliers (Stoltzfus, 2011). This assumption could be problematic given the presence of 'superforecasters' in the GJP dataset, who exhibit exceptionally high accuracy, and whose forecasts should be given higher weight, rather than transformed or removed to meet this assumption (Atanasov & Himmelstein, 2023). Furthermore, logistic regression assumes that there are linear relationships between the predictor variables and the log odds of the outcome (Stoltzfus, 2011), which may not hold true in a large dataset with a wider range of questions and participants than Shinitzky et al. (2024). This study compares the performance of BNs to regularized logistic regression, which was the highest performing machine learning method explored by Shinitzky et al. (2024).

The current study also incorporates forecast aggregation and recalibration. Shinitzky et al. (2024) used the absolute distance from a consensus forecast as their baseline predictor, where the consensus was the unweighted average of forecast probabilities for a question. The current study uses the information available in the GJP dataset to create a consensus aggregate that applies recalibration to an aggregate weighted by participant accuracy, forecast update frequency, and temporal proximity. This follows Atanasov and Himmelstein's (2023) procedure for calculating the consensus forecast for their proxy score. The absolute difference from the recalibrated aggregate is used as the baseline predictor in the current study. The consensus aggregate is also treated as a separate predictive model to serve as a basis of comparison for the other models, since this form of recalibration and aggregation can produce highly accurate forecasts (Atanasov et al., 2017).

Given that the recalibrated weighted aggregate is used as a component of the baseline predictor in all models, the first hypothesis of this study is that as more variables are included in the models, the BN with complex dependencies will have higher performance than the recalibrated weighted aggregate alone. The second hypothesis is that both structure-learned and flat BNs will have a higher performance than the equivalent logistic regression models, since BNs make fewer parametric assumptions than logistic regression, some of which are likely violated in a dataset like the GJP. The third hypothesis is that as the models include more variables, structure-learned BNs will have higher performance than flat BNs, since capturing more complex conditional dependencies will likely increase the information available for predicting the outcome.

## 2. Materials and Methods

To describe the methodology of the data analysis, I first characterize the dataset and the transformations applied to derive the continuous predictor variables. Subsequently, I outline the discretization process for the continuous variables, followed by the procedures employed to train the BN and logistic regression models. Finally, I detail the evaluation protocols applied to all models.

The dataset consisted of binary, non-conditional and non-voided questions from all four years of the GJP ACE tournament, yielding 694,442 forecasts made by 9480 forecasters on 303 questions. An example question is "Will Serbia be officially granted EU candidacy by 31 December 2011?". Each forecast consisted of a probability value (also called 'forecast value') associated with the first binary option of the question and included a timestamp of when the forecast was made and when the question closed. Four continuous predictor variables (described in detail below) were derived from the dataset which served as the basis of models predicting the binary outcome of the question (0 or 1) on a per-forecast basis. Depending on the model type, the continuous predictor variables were also discretized following multiple discretization configurations of each variable (described further below). For the sake of clarity, the models are grouped into tiers which correspond to the number of variables included as predictors. Each subsequent tier includes the predictors of previous tiers. Details of the predictor variables are described next according to the tier of model they first appeared in, followed by the discretization process.

The first tier of models included only the absolute difference of the forecast value from a consensus prediction value at the time of the prediction, similar to Shinitzky et al.'s (2024)'s baseline model. First, forecast values of .00 and 1.00 were replaced with .001 and .999, respectively, to avoid distortions in logarithmic operations using 0 (see Satopää et al. (2014) for a similar procedure). All predictors used this corrected forecast value, including the consensus prediction, described next.

The consensus prediction was a recalibrated weighted aggregate forecast, following the procedure described in Atanasov et al. (2017) and Atanasov and Himmelstein (2023). The weighted aggregate is described in (1),

$$\bar{p}_{t,k,l} = \left( \frac{1}{\sum_{t,i} \{d_{t,i,l} \times w_{t,i,l}\}} \right) \times \sum_{t,i} \{d_{t,i,l} \times w_{t,i,l}\} \times p_{t,i,k,l} \tag{1}$$

where $\bar{p}_{t,k,l}$ is the aggregate across forecasters, $i$, that depends on time, $t$, question, $l$, and outcome, $k$. The decay value, $d_{t,i,l}$, was set to 1 for 72% of the most recent forecasts within a question, and 0 otherwise. The weight value, $w_{t,i,l}$, depended on the accuracy score of a forecaster, $c$, and the update frequency score for a forecaster within a question, $f$ (see (2)).

$$w_{t,i,l} = c_{t,i}^{\gamma} \times f_{t,i,l}^{\delta} \qquad (2)$$

The accuracy score was based on the cumulative mean standardized Brier score (MSBS) for all questions that had resolved by the time the forecast was made (standardization was within-question). The frequency score was based on the number of forecasts made on the question up to the point of the current forecast. Both the accuracy and frequency score were rescaled to the range [0.1 to 1.0], where the least accurate / least frequent updater would receive 10% of the weight of the most accurate / most frequent updater. Both weight exponents, $\gamma$, and $\delta$, were held at 1.0. The accuracy weight was omitted when no questions had been resolved for a participant at the time of the forecast. The weighted aggregate was then recalibrated following (3),

$$\hat{p}_{t,k,l=} \frac{\bar{p}_{t,k,l}^{a}}{\bar{p}_{t,k,l}^{a} + \left(1 - \bar{p}_{t,k,l}\right)^{a}} \qquad (3)$$

where the extremizing constant of $a$ was 1.5.

The weighting and recalibration parameters were the same as those in Atanasov and Himmelstein (2023) and were not optimized for the dataset, so the predictive power of the aggregate forecast should be considered conservative. This recalibrated weighted aggregate was calculated for each forecast, such that it reflected the aggregate at the time the forecast was made.

The second tier of models included the forecast value as an additional predictor. The third tier of models included the number of days prior to the close of the question the forecast was made (henceforth 'days prior'). The fourth tier of models included the MSBS for a user at the time the forecast was made. For this tier of models, forecasts made by a participant prior to a single resolved question were discarded to avoid temporal leakage of accuracy information, yielding 619, 873 valid forecasts.

The MSBS was calculated by first calculating the binary Brier score of each forecast, i.e., the squared difference between the forecast value and the outcome (0 or 1). The forecast Brier scores were then standardized within each question, where the standardized value was the difference of each forecast value from the mean of the most recent forecasts made by other participants on the question, divided by their population standard deviation. If the standard deviation within a question was zero, the standardized score of a forecast was set to zero, to indicate no difference from the mean. Finally, the standardized Brier scores were averaged across questions that had resolved at the time of the current forecast, such that the MSBS reflected the average Brier score for a participant at the time of the current forecast. This score was then standardized and winsorized to be between -3 and 3 standard deviations of the mean due to the presence of some extreme outliers that negatively impacted the discretization process. It appears that extreme outliers were usually caused by early clustering of forecast values around a single mean value, leading to a small standard deviation, which would yield extremely large standardized scores for the small number of forecasts that were highly divergent from the mean. Given that these outliers appeared to legitimately reflect a large difference from the mean, but their exact value did not validly reflect the extremeness of the scores, winsorization was deemed an appropriate transformation. The discretization process also rendered the exact values of the outliers largely moot, as all extreme scores would tend to be binned into one discrete level.

For each tier of model, a set of discretization configurations for the variables were tested. For each configuration of discretized variables, three models were trained: a structure-learned BN, a flat BN, and a regularized logistic regression model using dummy variables to represent the discrete levels. A regularized logistic regression model was also trained on the continuous versions of the variables included in each tier of model. The details of each of these methodologies are described after explaining the discretization process. Finally, the raw recalibrated weighted aggregate forecast described in (3) was used as a separate baseline model to which the other models were compared, where the aggregate forecast value was treated as the classification probability.

## 2.1. Discretizing the Variables

The continuous variables described above were discretized using the Hartemink algorithm, as implemented in bnlearn version 5.1 (Hartemink, 2001; Scutari, 2010; Scutari & Silander, 2025). The algorithm attempts to retain as much total pairwise mutual information as possible while determining where to make discrete breaks in continuous variables (where total mutual information is the sum of mutual information between each pair of variables). The algorithm begins by discretizing the continuous variables into many levels (in this study, 120 levels of equal intervals). It then iteratively selects two neighboring levels in each variable to coalesce such that the least amount of total mutual information is lost in the coalescence. At each iteration, the number of discrete levels per variable decreases by one until it reaches either a pre-specified number, or a minimum of 1, where total mutual information = 0.

While choosing the precise number of discrete levels to retain for each variable requires some subjective judgement, the choices were informed by the information presented in Figure 1. The figure depicts the pairwise mutual information of each variable when all variables were discretized into the same number of levels, starting from 120 levels of equal intervals. A few trends can be observed. First, some variables contained much more pairwise information than others, with forecast value and difference from the aggregate containing roughly 43% and 32% of the information, respectively. In comparison, days prior and MSBS contained only about 8% and 4%, respectively. Second, the information decreased roughly logarithmically with each fewer level, with some variables showing a sharper decline as they approached the minimum of 2 levels. Notably, forecast value showed a sharp decline below 6 levels, and difference from the aggregate showed a sharp decline below 4 levels. Friedman et al. (2018) suggest that the average number of discrete bins required to avoid significant declines in precision for the GJP competition was 6 (11 for superforecasters), which additionally informed how many discrete levels the forecast value variable was given.

**Figure 1**

*Total mutual information as a function of the number of discrete levels*

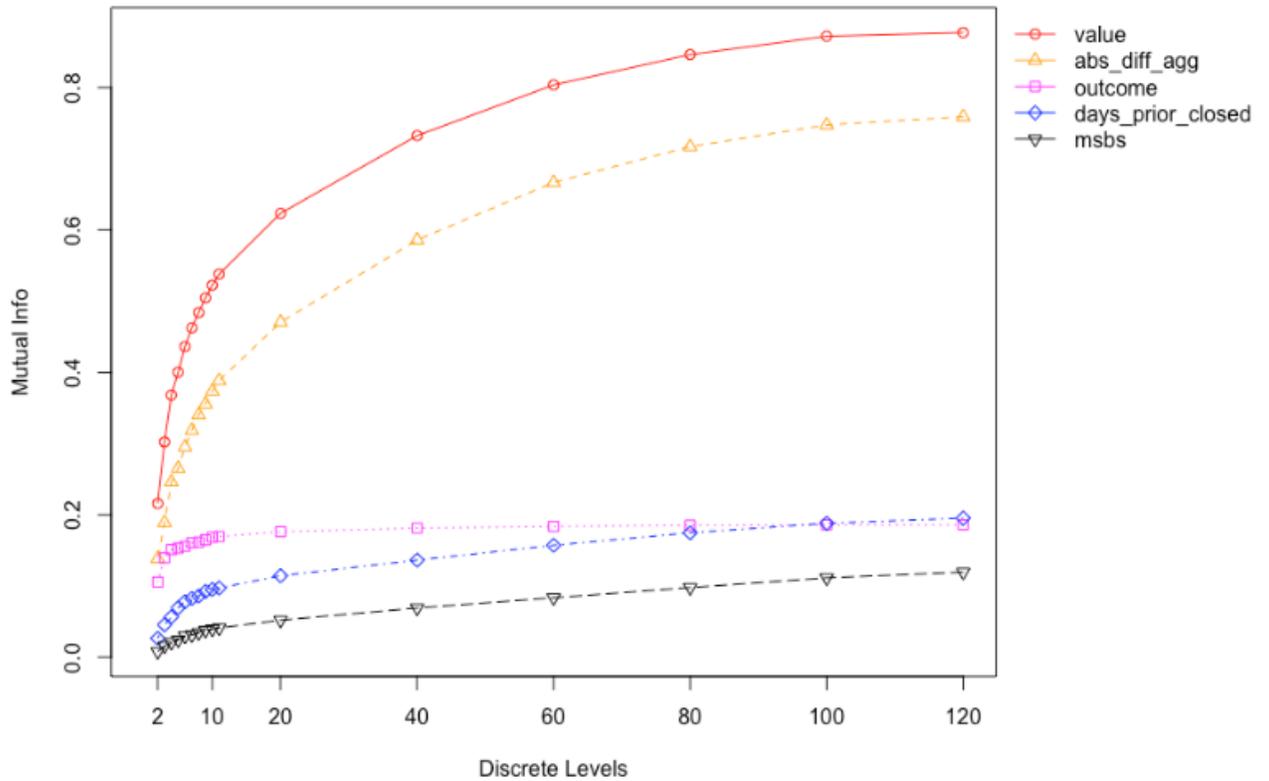

*Note*. Abs Diff Agg = Absolute Difference from the Aggregate, MSBS = Mean Standardized Brier Score.

These observations resulted in a few heuristics that were flexibly applied in deciding how many discrete levels each variable would receive. The first heuristic was that more levels should be given to variables that contained more mutual information. The second heuristic was that forecast value would have a target of between 11 and 6 bins, and absolute difference from the aggregate was given a target of at least 4.

As discussed by Hartemink (2001), there exists a trade-off between using more discrete levels per variable, which retains more mutual information, and the complexity of the resulting model, which can yield more parent-child relationships and thus more cells in the conditional probability tables of the BNs that must be estimated. More cells lead to fewer observations per cell, and nodes that are rarely or never observed cannot be estimated with certainty. Prior probabilities can be used to estimate unobserved combinations, as was done in this study, but in the absence of strong theoretical underpinnings of the prior specifications, overreliance on priors risks a poorly specified model.

To avoid over-reliance on priors in the BN parameterization caused by rarely observed parent-child configurations, discretization configurations were subject to the hard constraint that no more than 10% of the combinations of all discrete levels in the configuration could be unobserved. Following these heuristics and constraints, a set of discretization configurations was created for each tier of models (included in Table 1).

## 2.2. Training the Bayesian Networks

Training a BN is a two-step process of first learning the structure of the conditional dependencies between variables, followed by fitting the conditional probability tables of the variables, given the structure and the data. Given that the models were evaluated using 10-fold cross-validation (see below for details), 10 separate BN structures were learned on the training portion of each fold of the cross-validation setup. Structures were learned using the PC-stable algorithm implemented in bnlearn with the mutual information test. Arcs from the outcome to predictor variables were forbidden to ensure the predictors were parents of the outcome. Directed connections between variables were added if they appeared in the learned structures of > 6/10 of the training folds. Undirected connections with > 6/10 support were added and manually directed based on theoretical considerations. The same structure was used in each fold during the parameter fitting and cross-validation phase. The parameters of both the structure-learned and flat BNs were fit using maximum Bayesian posterior estimation, with a uniform prior corresponding to the Bayesian Dirichlet equivalent score (Scutari, 2025). Probability predictions about the outcome were generated from Bayesian likelihood weighting using the default of 500 random samples.

## 2.3. Logistic Regression Models

Logistic regression is a statistical methodology that models the probability of an outcome variable ($\hat{Y}$) being in one category or another (1 or 0) using a linear regression equation where each predictor variable ($X_i$) is weighted by its respective β coefficient. The β coefficients represent the slope of the regression line (which quantifies how much the outcome variable changes with a one-unit change in the predictor variable) and reveal the unique contribution of each predictor variable, after adjusting for the others (Stoltzfus, 2011). Regularization techniques apply a penalty to the β coefficients according to a tuning parameter, λ, which specifies the strength of the penalty. Regularization techniques are useful for logistic regression models with many potentially correlated predictor variables. Lasso/L1 regularization tends to pick one correlated coefficient and shrink the rest to zero, whereas ridge/L2 regularization tends to shrink correlated coefficients towards each other (Friedman et al., 2010).

The regularized logistic regression models in this study were fit with the glmnet R package version 4.1.10 (Friedman et al., 2010, 2025). For cross-validation, glmnet uses a grid search to find the value of λ that minimizes the mean loss across the 10 folds. This λ is then used in all 10 folds of the subsequent evaluation of the model. Following Shinitzky et al. (2024), the Tier 1 discretized model was fit with ridge/L2 regularization. Lasso/L1 regularization was used for all other tiers. Because the Tier 1 continuous model used only a single predictor, regularization using glmnet was inappropriate, so non-regularized logistic regression was performed using the glm function from the stats package in R (version 4.5.0).

## 2.4. Model Evaluation

Ten-fold cross validation was used to evaluate the models. The folds were balanced across outcome, with questions being segregated across folds to avoid information leakage. The same folds were used for each model within a discretization configuration, whereas folds were randomly generated across configurations. All models were evaluated using AUC as calculated by the pROC R package, version 1.19.0.1 (Robin et al., 2011).

## 3. Results

The AUC results suggest that the baseline recalibrated aggregate forecast was superior to both the Bayesian and logistic regression approaches, contrary to the first hypothesis of this study (see Table 1 for AUC results for all models). Both Bayesian approaches showed higher AUC scores than both logistic regression approaches, consistent with the second hypothesis of the study. However, the structure-learned BNs showed equivalent AUC scores to the flat BNs, contrary to the third hypothesis of the study. The discretized and continuous regularized logistic regression approaches both showed roughly equivalent AUC scores across all model configurations.

**Table 1**

*Performance Metrics for Baseline, Bayesian Network, and Logistic Regression Models*

| Config | No. discrete levels per variable | | | | | | Mean (SD) AUC | | | | BIC | |
| | Abs diff agg | Fcast Value | Days Prior | MSBS | BN | Flat BN | Discrete LogReg | Continuous LogReg | Aggregate | BN | Flat BN |
|---|---|---|---|---|---|---|---|---|---|---|---|
| Baseline | | | | | | | | | .985 (.018) | | |
| Tier 1 | | | | | | | | | | | |
| 1.0 | 80 | 0 | 0 | 0 | .602 (.047) | .602 (.047) | .606 (.047) | .580 (.051) | | -3013712 | -3013712 |
| Tier 2 | | | | | | | | | | | |
| 2.0 | 4 | 10 | 0 | 0 | .939 (.056) | .939 (.056) | .830 (.071) | .825 (.074) | | -2334986 | -2501996 |
| 2.1 | 5 | 8 | 0 | 0 | .938 (.041) | .938 (.041) | .831 (.056) | .824 (.061) | | -2444450 | -2626347 |
| 2.2 | 5 | 7 | 0 | 0 | .939 (.036) | .939 (.036) | .831 (.050) | .826 (.046) | | -2348550 | -2525157 |
| 2.3 | 4 | 6 | 0 | 0 | **.940 (.036)** | **.940 (.036)** | **.839 (.060)** | **.834 (.060)** | | **-1975226** | **-2128432** |
| Tier 3 | | | | | | | | | | | |
| 3.0 | 3 | 6 | 3 | 0 | .925 (.046) | .925 (.045) | .827 (.064) | .814 (.059) | | -2613377 | -2785402 |
| 3.1 | 3 | 5 | 3 | 0 | .932 (.039) | .932 (.039) | .829 (.040) | .829 (.032) | | -2455466 | -2622747 |
| 3.2 | 3 | 5 | 2 | 0 | .929 (.049) | .929 (.049) | .817 (.078) | .829 (.072) | | -2143187 | -2300066 |
| Tier 4 | | | | | | | | | | | |
| 4.0 | 3 | 4 | 2 | 2 | .929 (.046) | .930 (.046) | .824 (.059) | .833 (.056) | | -2157799 | -2297388 |

*Note.* Abs diff agg = Absolute Difference from Aggregate, AUC = Area Under the (receiver operating characteristic) Curve, BIC = Bayesian Information Criterion, BN = Bayesian Network (structure-learned), Fcast = Forecast, Flat BN = Flat Bayesian Network, LogReg = Logistic Regression, MSBS = Mean Standardized Brier Score.

For the Bayesian and logistic regression models, the highest AUC values were found when the models included only the forecast value and the absolute difference from the aggregate. Including days prior and MSBS did not improve AUC scores. Interestingly, within the best-performing tier of models (Tier 2), the number of discrete levels for both variables did not appear to have much effect on AUC. Descriptively, the highest performing model had the fewest discrete levels for forecast value and absolute difference from the aggregate (six and four levels, respectively). It thus appears that six discrete levels for forecast value was sufficient to avoid errors due to imprecision, as might be suggested by Friedman et al. (2018).

To further investigate why the structure-learned and flat BNs did not differ in AUC, the Bayesian Information Criterion (BIC) score of the models using the entire dataset were recorded (see final two columns of Table 1). The BIC score measures the log-likelihood of the data given the model, minus a penalty for model complexity (Kitson et al., 2023). This can serve as an indicator of whether any gains in explanatory power were worth the cost of a more complex model (e.g. with more structure / conditional dependencies). More positive values (closer to zero) represent a more likely and/or more parsimonious model. The structure-learned BNs had consistently more positive BIC scores than flat BNs, indicating that there were gains in explanatory power that overcame the penalty for greater complexity. However, it appears these gains were not large enough to result in meaningfully more predictive models. The most positive BIC score belonged to the configuration with the highest AUC and included only forecast value and absolute difference from the aggregate, with the fewest discrete levels each.

## 3.1. Relationships Between Predictor Variables

Examining the arrangement of the structure-learned BN models can show how the variables related to one another and can reveal information about the efficacy of the model. The structure-learning algorithm found that the outcome variable was conditionally dependent on absolute difference from the aggregate, forecast value, and days prior, for all models that included these variables. The Tier 4 model found that MSBS was conditionally independent of outcome, given the other predictor variables (see Figure 2). These results suggest that MSBS did not provide unique information related to the outcome beyond what was already present in the other variables. Overall, the structures of the BNs were consistent across models, and resembled that of Figure 2 (potentially with fewer variables, depending on the model tier).

**Figure 2**

*The Bayesian network structure from Tier 4*

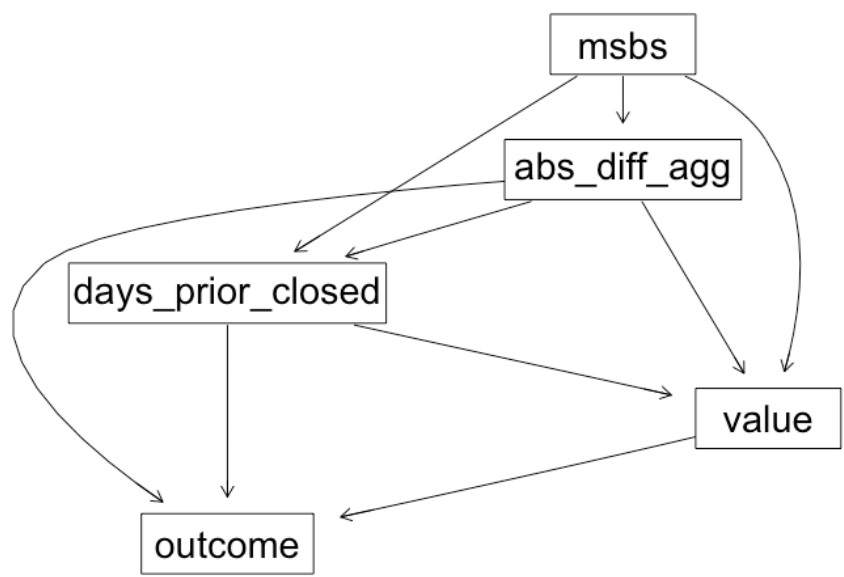

*Note*. Abs Diff Agg = Absolute Difference from the Aggregate, MSBS = Mean Standardized Brier Score.

In a similar vein, examining the coefficients from the logistic regression models can shed light on the how the variables were related and the efficacy of the models. The coefficients from the Tier 4 continuous logistic regression model suggested that MSBS and absolute difference from the aggregate contributed little unique predictive power, as their coefficients shrank to zero from the lasso regularization (see Table 2). Forecast value was the variable with by far the strongest relationship to the outcome. These results also suggest that the predictor variables contained redundant information for predicting the outcome.

**Table 2**

*Coefficients from the Tier 4 Lasso Logistic Regression Model with Continuous Variables*

| Predictor | Coefficient |
|---|---|
| (Intercept) | -2.33 |
| Forecast Value | 2.61 |
| Days Prior Closed | -0.0003 |
| Abs Diff Agg | - |
| MSBS | - |

*Note.* Abs Diff Agg = Absolute Difference from the Aggregate, MSBS = Mean Standardized Brier Score.

The results from the discretized Tier 4 logistic regression model are more complex but suggest that the most influential variables were forecast value and large values of absolute difference from the aggregate (see Table A1). MSBS and long-range forecasts received relatively little weight.

**3.2. Testing the Assumptions of Logistic Regression**

Examining the assumptions of the logistic regression models can illuminate why they had lower AUC values than the Bayesian models. The following five assumptions of logistic regression were tested on a non-regularized version of the Tier 4 continuous logistic regression model using the full dataset: independence of errors, linearity of predictor variables and the log-odds of the outcome, lack of multicollinearity among predictors, a lack of influential outliers, and a sufficiently large sample size (Stoltzfus, 2011). The assumption of independence of errors was violated on a priori grounds, since participants could provide repeated forecasts on the same question, leading to non-independence due to participant and question. Violating this assumption tends to artificially reduce the variance in parameter estimates, yielding overly small confidence intervals and $p$ values, and can also bias parameter estimates (Ives & Zhu, 2006; Roberts et al., 2017). It should be noted that BNs also typically assume independence of errors, so this is a shortcoming that applies to both types of models tested in this study (Azzimonti et al., 2022).

The assumption of a linear relationship between the predictors and the log-odds of the outcome and was tested with the Box-Tidwell test, where a logistic regression model was fit to the original predictors and interaction terms consisting of those predictors multiplied by their natural logarithms. All interaction terms were statistically significant ($p < .001$) and had large absolute $z$ values ($> 17$), indicating a violation of the assumption of linearity. Likewise, visual inspection of scatterplots of predictors versus the log odds of the outcome suggested a violation of linearity in the absolute difference from aggregate variable, as it displays a curvilinear U-shaped pattern (see Figure 3). Note that this assumption only applies to continuous variables and not dummy-coded variables, so it was not violated in the dichotomous logistic regression models (Stoltzfus, 2011).

**Figure 3**

*Scatterplot of Absolute Difference from Aggregate and Log Odds of Outcome*

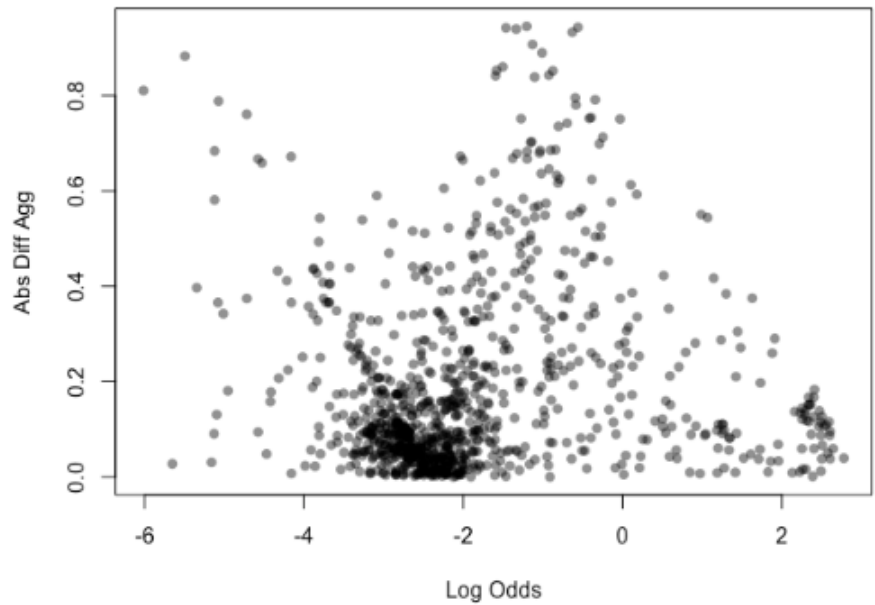

*Note.* To aid visualisation, points are from a random subset of n = 1000 data points. Abs Diff Agg = Absolute Difference from the Aggregate.

The assumption of the absence of influential outliers was tested by examining Cook's distances and standardized residuals. The results suggested that approximately 6% of the data had a Cook's distance > 4/*n*, and approximately 1% had an absolute standardized residual > 3, indicating a mild violation of this assumption. The assumption of the absence of multicollinearity was tested by examining the variance inflation factors (VIF) of the predictor variables. All VIF values were < 5, indicating no violation of this assumption. Finally, the assumption of a sufficiently large sample size was met, since a common rule of thumb is that there should be at least 10 observations of the least common outcome per independent variable, and the current sample consisted of > 600,000 observations with the least common outcome being approximately 20% (Stoltzfus, 2011). In sum, it appears that some of the assumptions of logistic regression were violated, particularly the assumptions of independence of errors and linearity. This likely led to somewhat poorly specified models, reducing their predictive ability relative to BNs which make fewer parametric assumptions.

## 4. Discussion

The results suggested that the recalibrated aggregate predictor alone was the most accurate at predicting outcome as measured by AUC, followed by the BN models, then regularized logistic regression models. Structure-learned BNs were not superior to flat BNs. Discretizing the variables did not have a meaningful effect on the performance of the logistic regression models.

There are a few plausible explanations for why BNs did not out-perform the recalibrated aggregate. The first is that the aggregate included information redundant with the other variables included in the models, namely, time information, forecast value, and accuracy. It was believed that absolute difference from the aggregate, conditional on the other variables, such as forecast value and MSBS, would provide a unique signal that would improve AUC over the raw aggregate. However, it seems that while conditional dependencies were found by the structure learning algorithm, they did not add enough additional information to increase AUC over flat networks. The fact that MSBS was not independently related to outcome, as found by the Tier 4 structure-learned BN, provides additional evidence that the variables contained redundant information related to participant accuracy.

Another plausible explanation for why the Bayesian methods did not outperform the raw aggregate is that the discretization process removed too much information from the continuous variables. As can be seen in Figure 1, reducing the number of discrete levels in forecast value from 120 to 10 results in a loss of approximately 40% of its total pairwise information. Likewise, the absolute difference from aggregate lost 65% of its total pairwise information between 120 and 5 levels (its highest level beyond Tier 1). Thus, the discretization process may have substantially reduced the information available to the models and thus lowered AUC scores.

The results highlight trade-offs in using discrete BNs to model realistic forecasting data. The upsides are that discrete BNs make few parametric assumptions and can model discrete outcomes directly (unlike typical hybrid BNs, where discrete nodes cannot be the child of continuous nodes). They are also transparent models of conditional dependencies between variables, which can be theoretically illuminating and sometimes highly related to model accuracy. Bayes nets are also able to incorporate theory-driven information in the form of expert-derived directional dependencies and prior conditional probabilities (Kitson et al., 2023).

The downsides are that the discretization process can substantially reduce the information available in continuous variables. The discretization process also requires a large dataset, as a greater number of discretized levels introduces more cells of conditional probability tables that must be estimated, reducing the number of observations available per cell. This puts a ceiling on the number of variables that can be included in a model. Finally, the discretization process requires some subjective judgement in deciding how many discrete levels each variable should have, with no guarantee of an optimized solution.

The relatively poor performance of the logistic regression models is likely owing to violations of the assumptions of the models, particularly independence of errors and linearity. Transformations of the variables to make them better fit the assumptions of logistic regression and forms of logistic regression better suited to repeated measures / correlated data would likely be an effective solutions to these issues (Stoltzfus, 2011). Non-standard hierarchical models that account for correlated data are available for both logistic regression and BN models, so exploration of these more advanced techniques is warranted, as are more complex blocking designs for the cross-validation procedure to account for dependencies in the data (Azzimonti et al., 2022; Ives & Zhu, 2006; Roberts et al., 2017). Additionally, explorations of interaction effects in the logistic regression models would also likely yield gains like those of conditional dependencies between nodes in a BN.

Future research wishing to expand on this study would first do well to examine a wider range of variables related to the outcome but not strongly related to the recalibrated aggregate. Atanasov and Himmelstein (2023) provide a set of variables worth exploring in this regard. Second, while discretizing all variables is required to model discrete outcomes directly using discrete BNs, a hybrid BN including both discretized and continuous variables could predict a continuous proxy of outcome (such as prediction error), which could then be used as a means of predicting outcomes. One example of this approach is similar to that suggested by Shinitzky et al. (2024), where the prediction probabilities of outcome from a logistic regression model were used as weights in a wisdom-of-the-crowds aggregation approach. Another related approach would be to use a continuous output of a BN as an input to a logistic regression model, which can be more effective than either method used alone (Wei & Dong, 2025). A hybrid BN could use discretized variables that do not fit the assumptions of linear or logistic regression and retain continuous variables that fit the assumptions of linear regression. A logistic regression model could then use this output, possibly in combination with other non-normal variables that fit the assumptions of logistic regression, to increase the predictive power of the ensemble.

**4.1. Conclusions**

Overall, the purpose of this study was to explore the utility of BNs in improving geopolitical forecasts, in comparison to logistic regression and recalibration and aggregation. Neither BNs nor logistic regression alone proved superior to state-of-the-art recalibration and aggregation methods under the conditions of this study. The results suggest that the findings of Shinitzky et al. (2024) are not robust to different datasets with different predictor variables, in that logistic regression was not the most accurate machine learning technique of the current study. While the current study did not improve on the accuracy of aggregation and recalibration techniques, the merits of BNs deserve further exploration. Investigations into the use of other predictor variables, data transformations, the effect of discretization, and more advanced forms of BNs (and logistic regression), such as those that can account for correlated data, would be fruitful areas of future research. Furthermore, using ensemble methods that play to the strengths of each methodology remains a promising next step for this research direction.

**Appendix**

**Table A1**

*Coefficients from the Tier 4 Lasso Logistic Regression Model with Dichotomized Variables*

| Predictor | Coefficient |
| --- | --- |
| (Intercept) | -2.20 |
| Days Prior Closed [0,100.65] | 0.91 |
| Days Prior Closed (100.65,549] | -0.00000000027 |
| Forecast Value [0.001,0.208917] | -1.25 |
| Forecast Value (0.208917,0.408517] | - |
| Forecast Value (0.408517,0.807717] | 1.43 |
| Forecast Value (0.807717,0.999] | 2.68 |
| Abs Diff Agg [2.58926e-05,0.115111] | - |
| Abs Diff Agg (0.115111,0.353503] | 0.072 |
| Abs Diff Agg (0.353503,0.986473] | -1.18 |
| MSBS [-3,0] | 0.014 |
| MSBS (0,3] | - |

*Note.* Abs Diff Agg = Absolute Difference from the Aggregate, MSBS = Mean Standardized Brier Score.

**Supplementary Materials**

Supplementary materials, including analysis code can be found in the following repository:

https://osf.io/v4hjz/

**Declaration of Generative AI and AI-assisted Technologies in the**

**Manuscript Preparation Process**

During the preparation of this work the author used Claude Sonnet 3.5 – 4.5 and Claude Opus 3.5 – 4.5 to generate some of the analysis code and help edit the manuscript. After using this tool/service, the author reviewed and edited the content as needed and takes full responsibility for the content of the published article.